# Formation of quasi-free and bubble positronium states in water and aqueous solutions

S.V. Stepanov[1,2,*], G. Duplâtre[3], V.M. Byakov[1,4], D.S. Zvezhinskiy[1] and V.S. Subrahmanyam[3,**]

[1] *Institute for Theoretical and Experimental Physics, B. Cheremushkinskaya, 25, 117218, Moscow, Russia*
[2] *National Research Nuclear University "MEPhI", Kashirskoye shosse 31, Moscow, 115409, Russia*
[3] *Institut Pluridisciplinaire Hubert Curien, CNRS/IN2P3, BP 28 67037 Strasbourg, Cedex 2, France*
[4] *D.Mendeleyev University of Chemical Technology of Russia, Miusskaya sq., 9, 125047, Moscow, Russia*

[*]email: stepanov@itep.ru
[**]on leave of absence from *the Department of Physics, Banaras Hindu University, Varanasi, India*

**Abstract:** It is shown that in aqueous solutions a positronium atom is first formed in the quasi-free state, and, after 50-100 ps, becomes localized in a nanobubble. Analysis of the annihilation spectra of $NaNO_3$ aqueous solutions shows that the hydrated electron is not involved in the positronium (Ps) formation.

Most usually, the positron annihilation lifetime (LT) spectra in liquids (below we consider aqueous solutions) are well described in terms of three exponentials (3-E analysis). It is believed that this fact indicates that 1) Ps formation is a fast process, its duration not exceeding 10 ps, the typical width of one channel of the time analyzer, and 2) the Ps atom is not involved in the non-homogeneous diffusion-controlled intratrack chemical reactions with radiolytic products. However, several facts plead against this conventional analysis, in particular:

1) The ratio of $I_3$ to $I_1$ (intensities of the ortho-Ps and para-Ps components) is not 3:1, as theoretically expected, but rather close to 2:1, Fig. 1 [1];

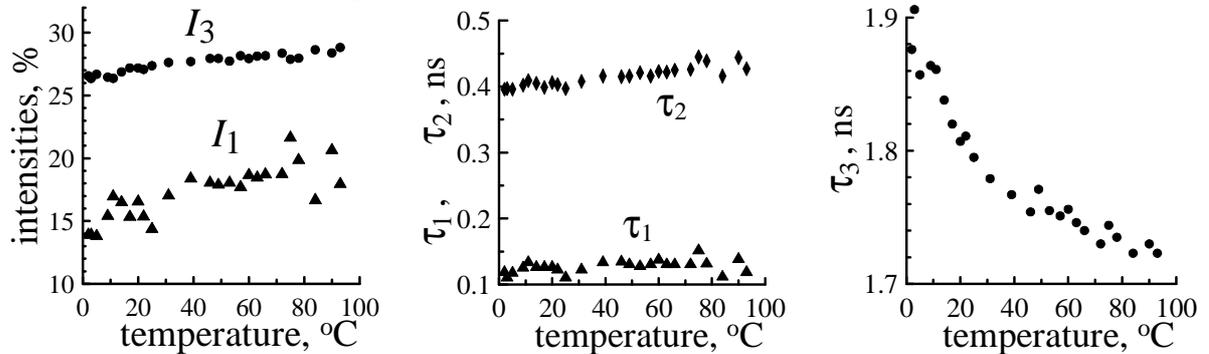

Fig. 1. Temperature dependence of the intensities of short-lived ($I_1$) and long-lived ($I_3$) components, and of the lifetimes $\tau_1$, $\tau_2$ and $\tau_3$ of LT spectrum of pure water [1].

2) The 3-E analysis cannot describe the behaviour of the *S*-parameter (characterizing the shape of the Doppler spectrum) with time at short times, the so-called "juvenile broadening" effect, Fig. 2 [2] (data on S(t) in pure $H_2O$ obtained by conventional AMOC technique is in [3]). This effect consists in a decrease of the *S*-parameter when we go from "para-Ps times" (~100-200 ps) to the negative times;

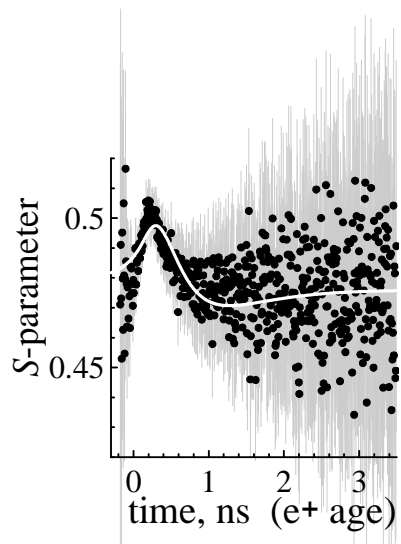

Fig. 2. Time dependence of the S-parameter in water at room temperature (recalculated from a GiPS-AMOC spectrum [2]). The solid line represents the fit of the AMOC spectrum, which takes into account appearance of the quasi-free positronium (see below). *S*-parameter is defined as the ratio of counts of annihilation photons within the energy 511±0.7 keV to that of 511±3 keV for each e$^+$ age.

3) The lifetime of the short-lived component of the LT spectrum, $\tau_1$, the para-positronium (p-Ps) lifetime in a 3-E fit, comes out to be less than the p-Ps lifetime in vacuum (124 ps). However, as it follows from the magnetic quenching experiments, the Ps contact density in water is 0.75, so that the p-Ps lifetime should be 150-160 ps [1];

4) The lifetime of the long-lived component, $\tau_3$, ascribed to ortho-positronium (o-Ps), decreases with increasing temperature (Fig. 1). This is inconsistent with the increase of the size of the Ps bubble ensuing from the decrease of the surface tension of water.

What hints give us these features of the simplest exponential treatment of the LT spectrum of pure water and how to interpret them?

Explanation of the anomalies of $\tau_1$ and $\tau_3$ lifetimes was already given in [1]. The low value of $\tau_1$ is simply an artifact in processing the LT spectrum. It is overcome by fixing $\tau_1$ to the value expected from magnetic quenching experiments and subsequent correction (very slight broadening) of the time resolution function of the LT spectrometer.

The decrease in $\tau_3$ with increasing *T* occurs due to the exponential increase of the reaction rate constant of the oxidation reactions of Ps by intratrack species (OH radicals, $H_3O^+$ ions). These reactions are diffusion-controlled and their rate constants increases exponentially with temperature. As a result, in spite of the drop in the pick-off annihilation rate (due to increase of the radius of the Ps bubble with increasing *T*), parameter $\tau_3$ decreases with temperature.

Discussion of the reasons for the first two anomalies (violation of the $I_3:I_1=3:1$ relationship and the maximum of *S*(*t*) at short times in water) as well as interpretation of the LT spectra of $NaNO_3$ aqueous solutions is the subject of this communication.

**General consideration**

The basic hypothesis we have used to interpret the experimental data is that localization of Ps (in a bubble) does not occur immediately. It has a relatively long-lived precursor (intermediate transient state): quasi-free positronium (qf-Ps). Confidence in this assumption about qf-Ps in water has become possible after measuring AMOC spectra with high statistics [2].

Qualitatively, the picture is as follows. After $^{22}$Na beta-plus decay, a fast positron (e$^{+*}$) enters a medium and loses energy through ionization. After thermalizing, it becomes solvated (or hydrated) similarly as does a quasi-free electron (the electron hydration time is 0.3 ps). All in all, it takes up to 10 ps [4]. The thermalized quasi-free (presolvated) positron can react with one of the quasi-free electrons in the terminal part of the e$^+$ track (e$^+$ blob). This is the essence of the recombination mechanism of Ps formation (also called the blob or spur model). As a result a quasi-free Ps, not yet localized in a bubble, is formed [5]. It is a loosely coupled (swollen) state of the $e^+e^-$ pair, located in a dielectric continuum undisturbed by the presence of the qf-Ps itself. The binding energy of qf-Ps in water is $\varepsilon^2 \approx 4$ times less ($\varepsilon \approx n^2 \approx 2$ is the high-frequency dielectric constant of water; n is the refractive index) than that in vacuum (approximately it is 1.7 eV, i.e. 5 eV less than in vacuum). In qf-Ps $e^+$ and $e^-$ are separated by a distance $\varepsilon$ times larger than in a vacuum. Consequently, the qf-Ps contact density in water is $\varepsilon^3 \approx 8$ times less [6]. This means that the annihilation rate of qf-Ps only slightly exceeds the annihilation rate of "free" $e^+$ (strictly speaking, hydrated $e^+$). For this reason we do not subdivide qf-Ps into ortho- and para-states, because both of them annihilate on outer molecular electrons within approximately the same time. The contribution to the $S(t)$-parameter from qf-Ps is approximately equal to that of free $e^+$, since the latter also annihilates with the same electrons.

Processing of the GiPS-AMOC spectrum of pure water at room $T$ [2] in terms of $S$-parameter clearly shows (Fig. 2) that the transformation of qf-Ps into the bubble state lasts $t_{loc}$ ~50-100 ps since the $e^+$ birth (in the bubble state because of p-Ps self-annihilation $S$-parameter gets larger value). This value far exceeds the duration of the Ps bubble growth (≤10 ps, derived on the basis of a solution of the Navier-Stokes equation) [7]. This fact implies that the bottle-neck of the formation of the Ps bubble state is not the growth of the bubble, but its initial stage – search for a preexisting trap (density fluctuation) by qf-Ps. It is well known that qf-Ps can not be trapped in a fairly small/shallow trap, because there is no bound state of Ps therein. The search for a deeper/larger trap requires longer time. This is the reason for some delay in formation of the equilibrium Ps bubble state [5].

The notion of a quasi-free Ps state, preceding the Ps bubble state, is not new. A similar approach was used in [8] to explain the juvenile broadening of the Doppler spectrum in some substances. The authors suggested that a "hot" positron knocks out an electron from a molecule in a medium and with a noticeable probability forms a hot (epithermal) Ps atom with a commensurate kinetic energy (10-25 eV). At a very short times (just after $e^+$ birth) this Ps kinetic energy broadens the Doppler spectrum (i.e., reduces the value of the $S$-parameter at $t \approx 0$). In [8] the authors assumed that the p-Ps contact density is equal to unity (as in vacuum), and hence adopted a p-Ps lifetime equal to that in vacuum. Finally, from their interpretation of the AMOC data, typical hot-Ps thermalization times (10-30 ps) were derived.

Our viewpoint is different. Firstly, we take it that the thermalization of subionizing $e^+$ and $e^-$ proceeds rather fast (fractions of ps), qf-Ps being formed from the thermalized particles [9]. The qf-Ps produced cannot have a kinetic energy much higher than its binding energy ($E_b$) in a dielectric continuum ($E_b \sim Ry/2\varepsilon^2 = 1.7$ eV), otherwise, it might just break up because of interaction of $e^+$ and $e^-$ with the environment, while qf-Ps moves through a medium.

Secondly, the qf-Ps lifetime is determined by the time needed to find of a suitable structural trap, able to capture and bind it (some qf-Ps annihilate on surrounding molecular electrons via the pick-off process). By fitting the GiPS-AMOC spectrum of pure water [2] and LT spectra of $NaNO_3$ aqueous solutions (see below) we have estimated the qf-Ps lifetime to be some 50-100 ps. Lower values of the S-parameter at $t \approx 0$ (juvenile broadening) are ascribed to qf-Ps (both its para- and ortho-states) annihilation.

Analysis of the LT spectra of pure water and $NaNO_3$ aqueous solutions at concentrations 0.07-0.31 M and various temperatures (measurements made by the Strasbourg group) was performed with the help of our program, which allows one to test various scenarios for Ps formation and kinetics of the subsequent intrablob reactions between radiolytic products and Ps until its annihilation. As the Ps precursors we have considered quasi-free (presolvated) and solvated electrons. We have also taken into account the oxidation reactions of the Ps localized in a bubble with OH radicals and $H_3O^+$ ions, and Ps ortho-para conversion by the radical species (OH radicals, hydrated electrons). The possibility for the epithermal positron to escape outside the blob at the final stage of its thermalization was also taken into account.

To describe accumulation of the main radiolytic products in water and Ps reactions in the $e^+$ blob we used non-homogeneous kinetic equations in the prescribed diffusion approximation [10]. Particularly, there are four equations related to populations of the $e^+$ states ("free" $e^+$, qf-Ps, o-Ps, p-Ps). These equations take into account Ps oxidation only (the contribution of the Ps ortho-para conversion is negligible):

"free" $e^+$: $dn_+/dt = -\lambda_+ n_+ + R_{oxi}(t)(n_o + n_p)$, $\quad n_+(t=0) = 1-P_{qf-Ps}$, (1)

qf-Ps: $dn_{qf-Ps}/dt = -(\lambda_{loc} + \lambda_{qf-Ps}) n_{qf-Ps}$, $\quad => \quad n_{qf-Ps}(t) = P_{qf-Ps} \exp[-(\lambda_{loc}+\lambda_{qf-Ps})t]$,

o-Ps: $dn_o/dt = 3\lambda_{loc} n_{qf-Ps}/4 - (R_{oxi}(t) + \lambda_{oPs}) n_o$, $\quad n_o(t=0) = 0$,

p-Ps:    $dn_p/dt = \lambda_{loc} n_{qf\text{-}Ps}/4 - (R_{oxi}(t) + \lambda_{pPs}) n_p, \quad n_p(t=0) = 0.$

The initial conditions for these equations refer to some 10 ps after the birth of $e^+$. By this time the formation of the $e^+$ blob and hydration of $e^-$ and $e^+$ are completed, and qf-Ps is formed (with probability $P_{qf\text{-}Ps}$). Accordingly, the probability of formation of the hydrated ("free") positrons is $1 - P_{qf\text{-}Ps}$. Each positron state (free $e^+$, qf-Ps, o-Ps, p-Ps) annihilates with its own rate constant: $\lambda_+$, $\lambda_{qf\text{-}Ps}$, $\lambda_{oPs}$, $\lambda_{pPs}$. To reduce the number of parameters we have assumed that $\lambda_+ = \lambda_{qf\text{-}Ps}$, and $\lambda_{loc} = 1/t_{loc}$ - transformation rate constant of qf-Ps in a bubble state. In the fitting procedure of the LT spectra $t_{loc}$ (average time of searching the trap and its growing to the equilibrium size) was set to to 50-100 ps, as induced from the analysis of the GiPS-AMOC spectrum [2]. At $t > t_{loc}$ qf-Ps is practically absent: partly annihilated, partly transformed into a bubble state (both ortho- and para-states, as expected, in a ratio of 3/1). $R_{oxi}(t)$ describes oxidation of Ps (in a bubble state), i.e. its reversion to a hydrated $e^+$, although complex formation may occur too (positron + negative ion of the oxidizer: the $e^+$ annihilation rate in this complex should be close to that of a free $e^+$). According to the model of $e^+$ blob, the value $R_{oxi}(t)$ has the form (see [9, 10]):

$$R_{oxi}(t) = \sum k_{i,Ps} c_i / V_{i,Ps}(t), \qquad V_{i,Ps}(t) = V_{bl} \cdot [1 + 4(D_i + D_{Ps})t/a_{bl}^2]^{3/2}, \qquad (2)$$
$$k_{i,Ps}(t) = k_{oxi} \cdot \left(1 + (R_i + R_{Ps})/\sqrt{\pi(D_i + D_{Ps})t}\right).$$

Here $k_{i,Ps}$ is the Ps oxidation rate constant by radiolytic oxidizers (OH, $H_3O^+$), $D_i + D_{Ps}$ is the sum of the diffusion coefficients of reactants, $a_{bl}$ is the initial size of the blob and $V_0 = (2\pi)^{3/2} a_{bl}^3$ is its characteristic volume. Substituting typical numbers ($a_{bl} \sim 40$ Å, $D_i + D_{Ps} \sim 5 \cdot 10^{-5}$ cm$^2$/s), we find that the characteristic time of the blob expansion is 0.5-1 ns: clearly, the intrablob oxidation of Ps implies inapplicability of the 3-E analysis of LT spectra. According to the latter approach, it is assumed that $I_1$ describes the decay of p-Ps, and $I_3$ - o-Ps. As follows from the above consideration $I_1$ includes also decays of qf-Ps, because they occur on a very short time-scale:

$$I_{qf\text{-}Ps} = \lambda_{qf\text{-}Ps} \int_0^\infty n_{qf\text{-}Ps} dt = P_{qf\text{-}Ps} \lambda_{qf\text{-}Ps} / (\lambda_{qf\text{-}Ps} + \lambda_{loc}). \qquad (3)$$

This contribution increases $I_1$ and, therefore, the $I_3/I_1$ ratio approaches to 2/1 which is in reasonable agreement with the experiment, Fig. 1.

**Ps formation in NaNO$_3$ aqueous solutions at different temperatures**

Radiation-chemical data suggest that the nitrate ion is an efficient scavenger of a "hot" electron (precursor of a thermalized electron). This is deduced from the exponential inhibition (suppression) of the yield of the hydrated electron vs. scavenger concentration (Fig. 3):

$G_e(c_S) \sim \exp(-c_S/c_{37}),$    for $NO_3^-$   $c_{37} = 0.53$ M.  (4)

Here $c_{37}$ is the concentration of the scavenger at which the yield decreeses 1/e (=0.37) times.

This dependence can be obtained on theoretical grounds, provided the electron capture (by a scavenger) occurs earlier than electron hydration, i.e. capture and hydration do not compete but are subsequent.

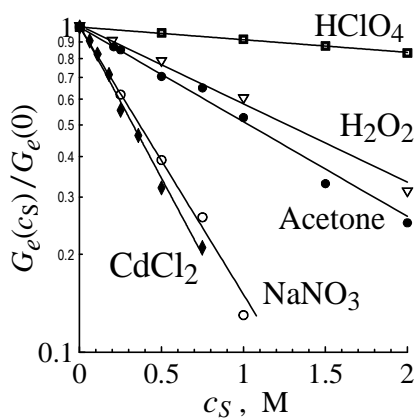

Fig. 3. Inhibition of the hydrated electron yield in aqueous solutions of various electron scavengers [11]

It is also known that $NO_3^-$ ions react with hydrated electrons (rate constant at about $k(NO_3^-) \approx 10^{10}$ $M^{-1}s^{-1}$). Therefore, being in a solution at concentration $[NO_3^-] \sim 0.1$ M, $NO_3^-$ ions during 1 ns may capture most of the hydrated electrons in the $e^+$ blob: $\exp(-k(NO_3^-)*[NO_3^-]*(1\ ns)) \approx 1/3$ (typical concentration of the hydrated electrons therein is ~0.01 M). However, processing of the LT spectra of nitrate aqueous solutions showed that Ps inhibition (more precisely, qf-Ps inhibition) takes place in agreement with the exponential law (Eq. 4) :

$$P_{qf-Ps}(c_S) = P_{qf-Ps}(c_S=0) \cdot \exp(-c_S/c_{37}) , \qquad (5)$$

where $c_{37}$ has the same numerical value (0.53 M) as in Eq. 4, describing inhibition of the hydrated electrons in picosecond pulse radiolysis experiments. Thus, capture of hydrated $e^-$ by $NO_3^-$ ions does not affect Ps formation probability. It ensues that the hydrated $e^-$ is not a Ps precursor; the (hydrated) positron does not react with it. So, in the case of nitrate solutions, in the system (1) just $P_{qf-Ps}(c_S/c_{37})$ must be used as the initial condition should as the initial probability of qf-Ps formation.

Some parameters obtained in fitting the LT spectra of $NaNO_3$ aqueous solutions are shown in Fig. 4. Therein $f_{oPs}$, $f_{qf-Ps}$ and $f_{pPs}$ are the proportions of $e^+$ annihilation in its different states (o-Ps, qf-Ps, p-Ps, respectively). To draw a rough analogy with 3-E analysis, we must adopt $I_3 <=> f_{oPs}$, $I_1 <=> f_{qf-Ps} + f_{pPs}$ and $P_{qf-Ps} <=> I_3 + I_1$ - total probability for Ps. So qf-Ps decays mostly contribute to $I_1$.

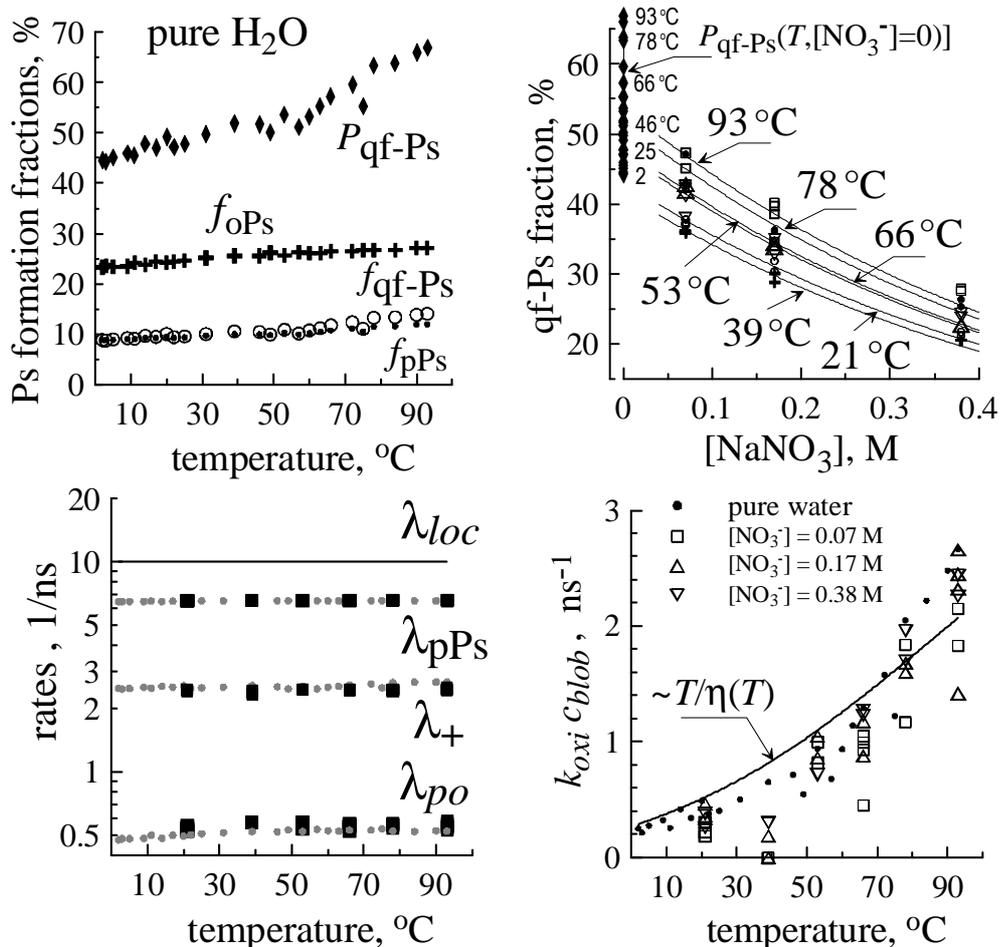

Fig. 4. Temperature and concentration dependences of the decay fractions (• - $f_{pPs}$ ; ○ - $f_{qf-Ps}$ and + - $f_{oPs}$) of different $e^+$ states (free $e^+$, qf-Ps, p-Ps, o-Ps), qf-Ps formation probability, $P_{qf-Ps}(T, c_S=[NO_3^-])$, the annihilation rates (localization rate constant is fixed at 10 $ns^{-1}$; bold symbols show data obtained in nitrate solutions, small symbols – in pure water) and the oxidation rate constant $k_{oxi}$ (solid line shows the Stokes-Einstein temperature behaviour of the diffusion controlled reaction rate constant; $\eta(T)$ is the viscosity of water).

**Conclusions**

1) Inclusion of quasi-free Ps, as a precursor of the Ps bubble state, into the Ps formation scheme allows to understand the cause of the underestimated ratio $I_3/I_1$ (~2/1 instead of 3/1). It also naturally explains the presence of a "hump" of the *S*-parameter at about 100 ps after the $e^+$ birth ("juvenile broadening"), attributable to the fact that annihilation of qf-Ps results in a wider Doppler spectrum than para-Ps, localized in a bubble.

A better understanding of the physical nature of qf-Ps comes from ACAR experiments in crystalline ice, where qf-Ps is observed in the Bloch state (formation of the bubble state is not possible there). For simplicity, on interpreting the LT and AMOC spectra we have assumed that the qf-Ps annihilation rate is equal to that of the free $e^+$ (more correctly, hydrated $e^+$). However, this is not fully true. In the para-state qf-Ps can annihilate into two gammas with its own electron. Although the probability of this process is low (the qf-Ps contact density is small), it is this annihilation channel that leads to the appearance of the Bloch peaks in the ACAR spectrum. Note that a large fraction of the Bloch para-qf-Ps can be produced from the ortho-qf-Ps due to its spin conversion into para-state at longer times. In the nowadays classical ACAR experiment [12] this was possible due to the presence of paramagnetic impurities in the ice (dissolved atmospheric $O_2$) stimulating ortho-para conversion.

2) Analysis of the LT spectra of aqueous solutions of $NaNO_3$ confirms that hydrated electrons do not take part in Ps formation. Most likely, this is due to the minute energy gain in this reaction between hydrated species: the energy of the Ps bubble state is only slightly below the sum of the energies of the hydrated $e^+$ and $e^-$, whereas significant rearrangement of the surrounding molecules is needed in the reaction. It would be interesting to study Ps formation in solution of other electron scavengers (acidic aqueous solutions) and in other solvents.

3) The temperature dependence of the Ps oxidation reaction rate constant by intrablob radiolytic species (OH radicals, $H_3O^+$ ions) is well described by the Stokes-Einstein law ($\sim T/\eta(T)$, $\eta$ is the viscosity of water), indicating that this reaction is diffusion-controlled. The value of the constant (for example, at room *T*) is in a good agreement with its theoretical estimate $k_{oxi} \cdot c_{blob} = 4\pi(D_i + D_{Ps})(R_i + R_{Ps}) \cdot ([OH] + [H_3O^+]) \approx 0.4$ ns$^{-1}$, Fig. 4 [13];

4) Ps ortho-para conversion on intrablob radical species is not witnessed, due to the fact that in water a spin-converter (paramagnetic particle) as OH radicals is primarily also a strong oxidizer.

5) In water, the mobility of positively charged $H_3O^+$ ions is twice as large as that of the hydrated electrons. As a result, an excess negative charge appears in the center of the blob which retains thermalized $e^+$ inside the blob. Debye screening of the $e^+$ charge by other charged blob species also leads to $e^+$ confinement inside the blob. In other molecular substances where electron mobility is greater than the mobility of positive ions, there is an opposite effect: $e^+$ may escape during thermalization outside the blob [14].

6) The Smoluchowski time correction (the factor $1 + (R_i + R_{Ps})/\sqrt{\pi(D_i + D_{Ps})t}$ in Eq. (2)) for the diffusion-controlled reaction rate constant is really negligible, since the diffusion displacement ($\sqrt{\pi(D_i + D_{Ps})t}$) of the reagents for typical times 0.1-1 ns is much larger than the reaction radius $R_i + R_{Ps}$ (a few angstroms). Only for $t \sim 1$ ps, $(R_i + R_{Ps})/\sqrt{\pi(D_i + D_{Ps})t} \sim 1$.

7) The temperature increase of the positronium formation probability $P_{qf-Ps}$ is related to a decrease in the energy needed for the production of one ion-electron pair (i.e. a decrease in the ionization potential of water). Hence it is a consequence of the increase of the total number of electron-ion pairs in the $e^+$ blob. The second plot in Fig. 4 shows some difference between the values of $P_{qf-Ps}(T)$ in pure water and that extrapolated to zero concentration in $NaNO_3$ solutions. It is possible that nitrate anions capture "hot" $e^+$ to a small extent. This effect has often a resonant character, the acceptor capturing a positron of a certain energy, corresponding to the maximum of the capture cross sections.

This work was supported by the Russian Foundation for Basic Research (grant 11-03-01066).